%% file: paperQuant.tex
\documentclass[12pt,a4paper]{article}
\pdfoutput=1        
\usepackage{amssymb,amsmath}
\usepackage[left=1.3in,right=1.3in,top=1.2in,bottom=1.2in]{geometry}
\usepackage{graphicx,color}  
\usepackage{dcolumn}  
\usepackage{bm}  
\usepackage{cancel}
\usepackage{titlesec}
\usepackage[font=singlespacing,font=small,labelfont=bf]{caption}
\usepackage{setspace}
\usepackage[brazilian]{babel}
\usepackage[utf8]{inputenc}
\usepackage[T1]{fontenc}
\usepackage{mathrsfs}
\usepackage{icomma}  
\usepackage{cite}
\usepackage{caption}

\titleformat{\section}
{\normalfont\bfseries\filcenter}
{\thesection.}{0.5em}{}

\titleformat{\subsection}
{\normalfont\itshape\filcenter}
{\thesubsubsection.}{0.5em}{}

\begin{document}
\onehalfspacing

\vspace{0.4in}

\large

\centerline{\textbf{Resolvendo a equação de Schr\"odinger através da}}
\centerline{\textbf{substituição direta da série de potências}}
\normalsize

\vspace{0.2in}

\setlength{\footnotesep}{0.75cm}
\centerline{Fábio E. R. Campolim\footnote{fabio.campolim@ufabc.edu.br}}

\small

\vspace{0.1in}

\centerline{\textit{Nanosciences and Advanced Materials Postgraduate Program}}
\centerline{\textit{Universidade Federal do ABC (UFABC)}}
\centerline{\textit{Av. dos Estados, 5001, 09.210-580 Santo Andr\'e, S\~ao Paulo, Brazil}}
\centerline{(25 de Agosto, 2014)}

\vspace{0.4in}

\normalsize

\begin{abstract}
\label{Resumo}

\vspace{24pt}

Este trabalho apresenta um método direto e de alta acurácia para 
resolver equações diferenciais ordinárias, em particular a equação de 
Schr\"odinger em uma dimensão, através da substituição direta de uma 
solução em série de potências para obter um sistema puramente algébrico 
contendo as relações de recorrência entre os coeficientes da série.
De posse dessas relações de recorrência, e conhecendo a forma exata da
equação diferencial, é possível construir uma rotina extremamente 
simples, usando somente operações aritméticas básicas em um processo de 
refinamento iterativo, para encontrar soluções de altíssima acurácia a um
custo muito reduzido de recursos de máquina.
No caso da equação de Schr\"odinger, os autovalores de energia podem ser
estimados por um método simples, e a estimativa pode ser facilmente 
refinada com o uso das relações de recorrência até que se alcance a 
tolerância especificada, o que permite encontrar funções de onda de alta 
acurácia mesmo para problemas com campos externos intensos e estados com 
número quântico alto.
Nesta monografia, o método é brevemente descrito e, em seguida, utilizado 
para resolver alguns problemas simples em mecânica quântica em uma 
dimensão.

\vspace{24pt}
\noindent
Palavras-chave: equação de Schr\"odinger, equações diferenciais, 
relações de recorrência, solução exata, série de potências.
\end{abstract}

\clearpage

\centerline{\textbf{Abstract}}
\vspace{24pt}

This work presents a direct and highly accurate method to solve ordinary
differential equations, in particular the Schr\"odinger equation in one
dimension, through the direct substitution of a power series solution to
obtain a purely algebraical system containing the recurrence relations
among the series coefficients.
With these recurrence relations at hand, and knowing the exact form of
the differential equation, it is possible to build an extremely simple
routine, using only basic arithmetic operations in an iterative refining
process to find solutions of very high accuracy at a very low cost of 
machine resources.
In the case of the Schr\"odinger equation the energy eigenvalues may be
estimated by a very simple method, an this estimate may be easily refined
using the recurrence relations until the specified tolerance has been 
reached, which allows one to find high accuracy wavefunctions even for
problems with intense external fields and states with very high quantum 
numbers.
In this monograph, the method is briefly described and then used to solve
some simple problems in quantum mechanics in one dimension.

\vspace{24pt}


\normalsize

\include{cap1}

\include{bibli}

\end{document}

%% file: cap1.tex
\section{Introdução}
\label{capIntro}

Desde que Erwin Schr\"odinger publicou sua famosa equação para a função 
de onda da mecânica quântica há 88 anos atrás \cite{SCHRODINGER}, uma 
miríade de métodos matemáticos foram desenvolvidos para solucionar essa 
equação para os mais diversos sistemas físicos envolvendo átomos e 
moléculas e, de fato, todas as ciências dos materiais encontram imensa 
utilidade em ferramentas que implementam esses métodos.
Não é à toa que uma grande quantidade de recursos é destinada atualmente 
à pesquisa nessa área de simulação computacional, pois hoje há uma 
verdadeira `corrida armamentista' mundial por métodos mais rápidos e 
acurados de solução de problemas de muitos corpos em mecânica quântica.
A empreitada rendeu, por exemplo, os prêmios Nobel de química para 
\cite{NOBEL} Robert S. Mulliken em 1966, Rudolph Marcus em 1992, Walter 
Kohn e John Pople em 1998 e, mais recentemente, o de 2013 a M. Karplus, 
M. Levitt e A. Warshel, sem falar em todas as patentes e produtos de 
software de altíssimo valor comercial.
É que o chamado experimento \textit{in silico} de sistemas atomísticos é 
de grande auxílio no desenvolvimento de medicamentos, combustíveis, novos 
materiais e até no estudo das origens da vida, entre tantos outros 
assuntos de grande importância \cite{NATUREMAT,NATURECHEM}.
Simulando o comportamento de átomos e moléculas em computadores somos 
capazes de reduzir imensamente o custo e o risco da pesquisa e também de 
ampliar as possibilidades de investigação como antes não era possível.
Capacidade esta que carrega a promessa de valiosas melhorias de vida para 
a sociedade em tantas e diversas áreas.

Esta monografia apresenta a solução da equação de Schr\"odinger através
da substituição direta de uma série de potências para obter um sistema 
de relações de recorrência entre os coeficientes da série que é puramente 
algébrico. 
A solução de qualquer equação diferencial ordinária é dada diretamente 
pela relação de recorrência mestre quando os coeficientes da equação são 
conhecidos e analíticos, de modo que, quando esses coeficientes são 
expandidos em série obtém-se um sistema simples e direto, que pode ser 
trivialmente implementado em um programa de computador e que nos dá a 
solução da equação diferencial com acurácia arbitrária, limitada somente 
pela precisão numérica das rotinas implementadas e pelos recursos de 
máquina disponíveis.
A implementação numérica envolve somente operações aritméticas simples 
para encontrar os coeficientes da solução, não necessitando qualquer 
procedimento adicional de integração, derivação ou inversão de matrizes 
e, portanto, é muito rápida mesmo quando essa solução possui uma 
quantidade muito grande de termos.
Para a equação de Schr\"odinger, a autoenergia da solução deve ser 
previamente estimada, e então as relações de recorrência podem ser 
utilizadas para refinar essa autoenergia, à qualquer acurácia que se 
queira.

Para demonstração do conceito, o método é aplicado a problemas simples 
de mecânica quântica envolvendo uma partícula em uma dimensão.
Importa mencionar que não é a intenção do presente trabalho formular e 
analisar o método no sentido matemático mais formal e rigoroso, e sim a 
aplicação prática do método para encontrar soluções de alta acurácia para 
a equação de Schr\"odinger, obviamente sem prejuízo das devidas 
caracterizações numéricas.

\section{A série de potências}
\label{secSolCompleta}

Um conceito muito importante que permeia diversas áreas da matemática 
como a aproximação de funções, a interpolação, a regressão, a solução 
de equações diferenciais e/ou integrais e até mesmo a aprendizagem 
supervisionada de máquina, é o teorema da aproximação de Weierstrass que 
diz que toda função real e contínua de uma variável e limitada em um 
intervalo fechado $[a,b]$ pode ser aproximada no intervalo de maneira 
uniforme e com acurácia arbitrária por um polinômio que, aliás, é único.
De modo mais geral, toda função contínua por partes no intervalo $[a,b]$ 
pode ser aproximada com acurácia arbitrária por uma combinação linear de
funções de um conjunto de base completo \cite{ARFKEN}. 
Na verdade, essa é a própria definição de conjunto completo: um dado 
conjunto de funções é dito completo se a norma uniforme do erro da 
aproximação tende a zero quando o número de funções de base na combinação 
linear tende a infinito.

Seja uma função aproximante $\phi$, real e de uma variável $x$, escrita 
em termos de um conjunto de funções de base, no presente caso um 
polinômio,
\noindent
\begin{equation}
    \phi_N(x) = \sum_{j=0}^{N} c_j \, x^j,
\label{eq:seriePot}
\end{equation}
onde os coeficientes $c_j$ são os parâmetros ajustáveis da solução e 
$N$ é o grau do polinômio, de modo que quando $N \to \infty$, a solução
se torna a série de potências.
Seja também uma função alvo $\psi$, real e de uma variável, contínua em 
$[a,b]$, que deve ser aproximada por $\phi$, então o erro pontual da 
aproximação $\varepsilon$ é uma função real dada por
\noindent
\begin{equation}
    \varepsilon_N(x) = \phi_N(x) - \psi(x).
\label{eq:erroApprox1D}
\end{equation}

Com essa definição podemos construir a norma $L^p$ da função erro,
\noindent
\begin{equation}
    \| \varepsilon_N \|_p = \left( \int\limits_a^b
        \left| \varepsilon_N(x) \right|^p
        \, \, \text{d} {x} \right) ^{1/p},
\label{eq:normPepsilon1Dp}
\end{equation}
e, em particular, quando $p=2$,
\noindent
\begin{equation}
    \| \varepsilon_N \|_2 = \left( \int\limits_a^b
        \bigr| \varepsilon_N(x) \bigr|^2
        \, \, \text{d} {x} \right) ^{1/2},
\label{eq:normPepsilon1Dp2}
\end{equation}
temos o erro quadrático médio, ou norma euclidiana do erro, e quando 
$p \to \infty$,
\noindent
\begin{equation}
    \| \varepsilon_N \|_{\infty} = \underset{x \in [a,b]}{\text{max}} \, 
                                 \bigr| \varepsilon_N(x) \bigr|,
\label{eq:normPepsilon1DpInf}
\end{equation}
temos a norma uniforme, notando que a convergência pela norma uniforme
é sempre mais forte que a convergência de qualquer norma com $p$ finito,
em particular, a convergência pela norma uniforme implica em convergência
pela média quadrática, contudo o contrário não é necessariamente verdade.
A definição de conjunto completo surge então da situação limite em que o
polinômio possui infinitos termos, e a norma uniforme do erro com 
respeito ao intervalo $[a,b]$ é exatamente igual a zero\footnote{
O mesmo resultado também vale para a norma $L^2$ com respeito a uma 
função peso $w(x)$ multiplicada ao integrando \cite{ARFKEN}.},
\noindent
\begin{equation}
    \underset{N \to \infty}{\text{lim}} \| \varepsilon_N \|_{\infty} = 0.
\label{eq:condConjCompleto}
\end{equation}

Na aproximação de funções, se a função alvo $\psi(x)$ é analítica, como 
muitas das funções de onda dos problemas de interesse em mecânica 
quântica, então $\psi(x)$ pode ser expandida em série de Taylor, por 
exemplo, em torno de $x=0$, chamada de série de Maclaurin, e os 
coeficientes da série de potências são dados simplesmente por 
$c_j = \psi^{(j)}(0) / j!$, onde $\psi^{(j)}(0)$ é a $j$-ésima derivada
da função alvo calculada na origem \cite{ARFKEN}.
Ainda mais, na solução de equações diferenciais sabemos, pelo teorema de 
Fuchs, que sempre se pode encontrar uma solução em série de potências 
para a equação diferencial, desde que a expansão seja em torno de um 
ponto ordinário ou no máximo uma singularidade regular, o que é sempre 
satisfeito em problemas de mecânica quântica \cite{ARFKEN}.

O ponto chave é que a função de onda de qualquer partícula física em um 
problema realista de mecânica quântica sempre estará quase toda 
confinada a uma determinada região limitada do espaço, decaindo a zero 
exponencialmente ou mais rápido desde ali até o infinito.
Por exemplo \cite{GRIFFITHS,COHEN}, uma partícula ligada a um poço 
quadrado de potencial tem função de onda oscilatória dentro do poço, e 
exponencial decrescente para além da borda do poço (seção 
\ref{secOscLinear}).
O mesmo ocorre com um elétron ligado a um átomo, a função de onda do 
elétron pode oscilar na região em torno do núcleo, mas sempre decai 
exponencialmente para além da borda do átomo.
Para uma partícula ligada em um potencial quadrático, pode-se argumentar, 
o decaimento da solução é 
gaussiano mas, em uma situação realista, o potencial quadrático nunca se 
estende até o infinito, após certa distância o aparato termina e, no 
espaço livre, o decaimento da função de onda é novamente exponencial.
Mesmo quando uma partícula é livre, ela não se estende de fato como uma 
onda plana até o infinito, há sempre um `pacote' de ondas definindo uma 
região de extensão principal da partícula, por exemplo, um envelope
gaussiano\footnote{
A densidade de probabilidade de uma partícula não é 
necessariamente um só  `pacote', na mecânica quântica uma mesma partícula 
pode estar distribuída em diversos pacotes distintos no espaço.
Por exemplo, uma partícula que tunela entre dois poços de potencial 
adjacentes, criando neles oscilações distintas, e portanto números 
quânticos distintos.
Outro exemplo é a conversão paramétrica, onde um fóton incidente em um 
cristal muito especial é convertido em dois pacotes de onda distintos 
viajando em direções distintas, espacialmente separados, mas que são 
quânticamente correlacionados \cite{TWOPHOTON} como um par EPR 
(Einstein-Podolsky-Rosen), ou seja, formam um estado de `dois-fóton' 
(\textit{two-photon state} em inglês), propositalmente violando a flexão 
de número no nome para indicar um par correlacionado.
Em todos os casos, a função de onda decai a zero no infinito 
exponencialmente ou mais rápido.}.
O decaimento exponencial da função de onda garante a mais importante 
condição de contorno dos problemas em mecânica quântica, que a função de
onda seja nula no infinito \cite{ARFKEN},
\noindent
\begin{equation}
    \underset{x \to \pm \infty}{\text{lim}} \psi(x) = 0.
\label{eq:limPsiInf}
\end{equation}

Essas características físicas da solução nos trazem dois benefícios 
imediatos na análise.
O primeiro é que o comportamento assintótico de decaimento exponencial da
função de onda garante que o raio de convergência da série de potências
seja sempre infinito em qualquer problema realista,
\noindent
\begin{equation}
    R^{-1} = \underset{j \to \infty}{\text{lim}} \dfrac{c_{j+1}}{c_j} = 
             \underset{j \to \infty}{\text{lim}} \dfrac{j!}{(j+1)!} = 0,
\label{eq:radiusConv}
\end{equation}
de modo que a série de potências sempre converge em todo o domínio da 
coordenada \cite{ARFKEN}, $-\infty < x < \infty$.
As funções de onda que aproximaremos são as soluções dos problemas 
arquetípicos em mecânica quântica, e são todas funções suaves, 
infinitamente diferenciáveis, que oscilam em uma determinada região 
espacial e depois passam suavemente a um decaimento exponencial para 
além dessa região.
O segundo benefício é que sempre podemos encontrar uma região espacial 
limitada, aqui denominada região da fração principal da partícula, que 
contenha a vasta maioria da densidade de probabilidade da função de onda, 
e que pode substituir o domínio em todos os cálculos pertinentes e para 
todos os efeitos práticos, nos permitindo trabalhar numericamente, como 
veremos na seção seguinte.

De qualquer maneira, a condição (\ref{eq:condConjCompleto}) nos diz que 
o erro da aproximação no intervalo tende a zero monotonicamente quando o 
número de termos do aproximante tende a infinito, de modo que, na prática, 
calcula-se termos até que uma determinada tolerância especificada seja 
atingida.
Assim, se pretendemos obter uma solução com um determinado critério de 
tolerância, então haverá sempre um número de termos $N$ mínimo que 
satisfaz esse critério.
O polinômio de menor grau $N_{\text{min}}$ que satisfaz a aproximação de 
norma $p$ no intervalo definido $[a,b]$ dentro da tolerância especificada 
$\tau$, $\| \varepsilon_{N_{\text{min}}} \|_{p} \leq \tau$, é por vezes 
denominado polinômio \textit{minimax} \cite{MINIMAX} ou, mais 
genericamente, aproximante \textit{minimax} para um conjunto completo 
qualquer.

Uma característica importante da série de Maclaurin é que os primeiros 
coeficientes da série tratam de aproximar a função na região próxima ao 
ponto $x=0$, enquanto que os coeficientes de índices mais altos ajustam 
regiões progressivamente mais distantes da origem\footnote{
É sempre uma 
boa idéia ter as distribuições de carga aproximadamente centralizadas na 
origem de modo a equalizar os erros dentro do intervalo.}.
Assim, o tamanho do intervalo em que a aproximação satisfaz ao critério 
de tolerância tende crescer monotonicamente quando se aumenta o número 
de termos da série, e os maiores erros tendem a se acumular nos extremos 
do intervalo.
Uma vantagem disso é que, sob determinadas circunstâncias, como no 
decaimento exponencial, o erro pontual máximo ocorrerá sempre em um dos 
dois extremos do intervalo, e então  a norma uniforme do erro
$\| \varepsilon_N \|_{\infty}$ é igual ao maior valor entre 
$| \varepsilon_N(a) |$ e $| \varepsilon_N(b) |$.

Importa notar também que um polinômio de grau $N$ finito só pode 
aproximar com um mínimo de qualidade um certo intervalo finito da função 
alvo e, para além desse intervalo, o polinômio se torna um péssimo 
aproximante, certamente divergindo ao infinito em ambos os lados, 
simplesmente porque o termo $x^N$ acaba dominando a soma.
Assim, o intervalo aproximante de máximo comprimento que, para um dado 
polinômio de grau $N$, atende à tolerância especificada será denominado
intervalo \textit{maximin}, sendo nulo se o polinômio não atende à 
tolerância especificada em qualquer intervalo porque seu grau é 
insuficiente.
Em outras palavras, o intervalo \textit{maximin} é o intervalo em que a 
aproximação pelo polinômio é válida dentro de um determinado critério de
tolerância.
Então, com a devida diligência, deve ser possível encontrar um polinômio
aproximante com um número mínimo de termos $N_{min}$ dentro de um 
intervalo máximo $[a,b]_{max}$, ou seja, a aproximação de um polinômio 
\textit{minimax} em um intervalo \textit{maximin}, que contenha toda a 
fração principal da densidade de probabilidade da partícula, e no qual 
todos os cálculos são, para todos os efeitos práticos, equivalentes aos 
cálculos feitos com a série infinita sobre todo o domínio da coordenada.
%

Uma consequência da ligação entre o número de termos do polinômio 
\textit{minimax} e o tamanho do intervalo \textit{maximin} é que 
podemos definir uma região de `domínio' para cada termo da série de 
Maclaurin,
que é simplesmente a região 
adicionada ao intervalo \textit{maximin} a cada incremento de N.
Quando, por exemplo, a série aproxima uma função seno é necessário um
polinômio de grau 7 para aproximar bem 1 período da onda, um de grau 17
para aproximar bem 2 períodos, um de grau 23 para aproximar bem 3 
períodos, um de grau 33 para aproximar bem 4 períodos, e assim por diante 
de modo que, próximo à origem, cada período exige em média 4 termos para 
ser descrito, ou seja, a região de `domínio' de cada termo é de 
aproximadamente $\pi/2$ em média.
Contudo, podemos notar que o polinômio de grau 87 já não aproxima bem os
11 períodos e, portanto, a região de `domínio' de cada termo da série do 
seno deve decrescer com $N$, de modo que cada período extra se torna cada 
vez mais difícil de aproximar.

Já no decaimento exponencial, a região de domínio de cada termo aparenta 
ser aproximadamente constante com $N$, cada termo aproxima por volta de 
$3/8$ em unidades da coordenada, e podemos então esperar em nossa vantagem 
que a fração principal da partícula sempre poderá ser aproximada por um 
número relativamente restrito de coeficientes.

Convenientemente, as soluções que queremos aproximar oscilam em uma certa 
região e depois apresentam decaimento exponencial nos extremos do domínio 
da coordenada, e então o tamanho do intervalo \textit{maximin} cresce 
linearmente com o grau $N$ do polinômio aproximante, e o limite onde esse 
intervalo vai a infinito é obtido de brinde quando se faz $N \to \infty$.
Desse modo, na prática basta simplesmente aumentar o número de termos do 
polinômio, calculando o intervalo \textit{maximin} a cada termo, até 
atingir o grau $N_{min}$ onde os dois critérios sejam simultaneamente 
atendidos, que a norma uniforme se torne inferior à tolerância 
$\| \varepsilon_{N_{\text{min}}} \|_{\infty} \leq \tau$, e que o 
intervalo \textit{maximin} contenha todo o intervalo da fração principal, 
de acordo com uma medida específica que será descrita à frente.

Uma maneira simples para calcular o intervalo \textit{maximin} de um 
polinômio com um determinado grau $N$ é determinar os pontos onde o erro
descola do zero nas regiões $x<0$ e $x>0$, ou seja, o primeiro local onde 
$|\varepsilon_N(x)| = \tau$ a partir do zero.
Por exemplo, para encontrar o ponto na região $x>0$ onde o erro descola
do zero, podemos usar uma rotina numérica para encontrar o zero de uma 
função como $F_N(x) = (\tau - |\varepsilon_N(x)|) \, \text{e}^{-x}$ 
usando a origem como estimativa inicial. 
O primeiro coeficiente da série, $c_0$, faz o erro na origem ir a zero, 
de modo que $\varepsilon_N(0)=0$ e $F_N(0)=\tau$, e o termo exponencial 
garante que $F'_N(0)<0$ e, portanto, garante que a rotina vá buscar o 
zero na região $x>0$, já que o termo $\tau - |\varepsilon_N(x)|$ é mal 
condicionado na origem.
A rotina deve então convergir rapidamente para o zero no extremo do 
intervalo $F_N(b)=0$, onde o erro é o máximo tolerado
$|\varepsilon_N(b)| = \tau$.
Do mesmo modo, o ponto onde o erro descola do zero na região $x<0$ pode 
ser encontrado buscando o zero de 
$(\tau - |\varepsilon_N(x)|) \, \text{e}^{+x}$, também usando a origem 
como estimativa inicial.

\section{Substituição direta na equação diferencial}
\label{secSubsDireta}

Seja uma equação diferencial ordinária de segundo grau e linear escrita 
na forma
\noindent
\begin{equation}
    \dfrac{\text{d}^2}{\text{d}x^2} \psi(x) + P(x) \dfrac{\text{d}}{\text{d}x} \psi(x) 
         + Q(x) \psi(x) + R(x) = 0,
\label{eq:eqDif1Dnhom}
\end{equation}
onde $\psi$ é a solução exata da equação (função alvo), e as funções 
$P(x)$, $Q(x)$ e $R(x)$ são reais, sendo que $R(x)$ é um termo de fonte 
que torna a equação homogênea quando nulo\footnote{
O método aqui 
apresentado também pode ser aplicado para resolver equações e sistemas 
de equações diferenciais não-homogêneas, não-lineares, de graus mais 
altos, e em um número maior de variáveis independentes, bastando para 
isso tomar os polinômios e medidas apropriados.}.
Com relação às equações homogêneas, sabe-se que possuem duas e somente 
duas soluções independentes \cite{ARFKEN}.
Ainda mais, se adicionalmente à equação homogênea são impostas condições 
de contorno, pode haver uma multiplicidade de soluções não-triviais para 
o problema, e dizemos então que o problema é um problema de autovalores.

Agora, substituindo o aproximante (\ref{eq:seriePot}) na equação 
diferencial (\ref{eq:eqDif1Dnhom}) temos que esta não é mais 
identicamente nula e devemos substituir o lado direito pelo erro pontual,
\noindent
\begin{equation}
    \dfrac{\text{d}^2}{\text{d}x^2} \phi_N(x) + P(x) \dfrac{\text{d}}{\text{d}x} 
          \phi_N(x) + Q(x) \phi_N(x) + R(x) = \varepsilon_N(x),
\label{eq:eqDif1DhomErr}
\end{equation}
de modo que (\ref{eq:condConjCompleto}) ainda vale, agora no contexto de 
que o aumento do número de termos do aproximante diminui monotonicamente 
o erro da solução da equação diferencial, tornando exata a solução por 
série no intervalo selecionado $[a,b]$ quando $N \to \infty$.

O uso da solução polinomial (\ref{eq:seriePot}) torna trivial o cálculo 
das derivadas primeira e segunda em (\ref{eq:eqDif1DhomErr}),
\noindent
\begin{equation}
    \dfrac{\text{d}}{\text{d}x} \phi_N(x) = \sum_{j=1}^{N} j \, c_j \, x^{j-1},
\label{eq:seriePotddx}
\end{equation}
\noindent
\begin{equation}
    \dfrac{\text{d}^2}{\text{d}x^2} \phi_N(x) = \sum_{j=2}^{N} j \, (j-1) \, c_j \, x^{j-2},
\label{eq:seriePotd2dx2}
\end{equation}
e, em particular, para a série de potências $N \to \infty$,
\noindent
\begin{equation}
    \phi_\infty(x) = \sum_{j=0}^{\infty} c_j \, x^j,
\label{eq:seriePotInf}
\end{equation}
\noindent
\begin{equation}
    \dfrac{\text{d}}{\text{d}x} \phi_\infty(x) = 
         \sum_{j=1}^{\infty} j \, c_j \, x^{j-1} = 
         \sum_{j=0}^{\infty} (j+1) \, c_{j+1} \, x^j,
\label{eq:seriePotddxInf}
\end{equation}
\noindent
\begin{equation}
    \dfrac{\text{d}^2}{\text{d}x^2} \phi_\infty(x) = 
         \sum_{j=2}^{\infty} j \, (j-1) \, c_j \, x^{j-2} =
         \sum_{j=0}^{\infty} (j+2) (j+1) \, c_{j+2} \, x^j,
\label{eq:seriePotd2dx2Inf}
\end{equation}
de modo que, substituindo as expressões (\ref{eq:seriePotInf}) a 
(\ref{eq:seriePotd2dx2Inf}) em (\ref{eq:eqDif1Dnhom}), a equação 
diferencial é transformada em uma equação puramente algébrica e exata 
que lhe é completamente equivalente\footnote{
A substituição direta da solução por série de potências guarda 
semelhança com o método de Frobenius utilizado em análise 
local\cite{BENDER}, mas aqui a aplicação é mais direta, pois não requer 
o uso de equações indiciais.
Além disso, ao contrário da solução de Hylleraas \cite{HYLLERAAS1,HYLLERAAS2}
que emprega um polinômio multiplicado por uma exponencial, começar 
diretamente com a série de potências torna muito mais simples o cálculo 
das integrais das normas e valores esperados, entre outras vantagens 
descritas adiante.}.
Ainda mais, as funções $P(x)$, $Q(x)$ e $R(x)$ podem ser expandidas em
séries de potência, deixando ao final uma equação algébrica dada 
puramente pela combinação linear das funções de base $x^j$.

Por exemplo, a equação de Schr\"odinger independente do tempo para uma 
partícula em uma dimensão tem a forma $P(x)=0$ e $R(x)=0$,
\noindent
\begin{equation}
    \dfrac{\text{d}^2}{\text{d}x^2} \psi(x) + Q(x) \psi(x) = 0.
\label{eq:eqSchrodinger1DQ}
\end{equation}
Se, além da solução, também a função $Q(x)$ é expandida em uma série de 
potências,
\noindent
\begin{equation}
    Q(x) = \sum_{j=0}^{\infty} Q_j \, x^j,
\label{eq:expansaoQ}
\end{equation}
onde $Q_n$ são os coeficientes de $Q(x)$, então a equação 
(\ref{eq:eqSchrodinger1DQ}) se torna simplesmente
\noindent
\begin{equation}
    \sum_{j=0}^{\infty} (j+2) (j+1) \, c_{j+2} \, x^j + 
         \sum_{j'=0}^{\infty} Q_{j'} \, x^{j'} \,  
         \sum_{j=0}^{\infty} c_j \, x^j = 0,
\label{eq:eqSchrodinger1DQexpand}
\end{equation}
que, recombinando os termos, fica
\noindent
\begin{equation}
    \sum_{j=0}^{\infty} (j+2) (j+1) \, c_{j+2} \, x^j + 
         \sum_{j=0}^{\infty} \left( \sum_{i=0}^{j} c_i Q_{j-i} \right) x^j = 0.
\label{eq:eqSchrodinger1DQexpand2}
\end{equation}

Agora, da análise da unicidade das séries de potência \cite{ARFKEN}, 
sabemos que os coeficientes de cada potência de $x$ do lado esquerdo de
(\ref{eq:eqSchrodinger1DQexpand2}) devem se anular individualmente, de 
modo que obtemos a relação de recorrência,
\noindent
\begin{equation}
    \boxed{c_{j+2} = - \dfrac{1}{(j+2)(j+1)} \sum_{i=0}^{j} c_i Q_{j-i}},
\label{eq:eqSchrodinger1DRelRec}
\end{equation}
que, dados os coeficientes da função $Q(x)$ arbitrária, e dados os 
coeficientes independentes $c_0$ e $c_1$, que chamaremos de sementes, 
determina todos os coeficientes da solução a partir de $c_2$.
As sementes $c_0$ e $c_1$ são escolhidas de modo a satisfazer condições 
de contorno, de simetria e/ou de normalização do problema.
A solução mais geral tem $c_0 \ne 0$ e $c_1 \ne 0$, mas é possível obter
as duas soluções independentes do problema, por exemplo, anulando cada 
semente individualmente.
Quando se faz $c_0 = 0$ e $c_1 \ne 0$ obtemos a solução que cruza o zero
no ponto zero, e fazendo $c_0 \ne 0$ e $c_1 = 0$ obtemos a outra 
solução\footnote{
Para equações diferenciais de grau maiores que o 
segundo, o número de sementes e, portanto, de soluções independentes,
é igual ao grau da equação, por exemplo, o termo de terceiro grau 
$d^3/dx^3$ leva a uma relação de recorrência para $c_{j+3}$, deixando 
independentes os coeficientes $c_0$, $c_1$ e $c_2$.}.

Para resolver uma equação diferencial homogênea mais geral, tomando 
$R(x)=0$ em (\ref{eq:eqDif1Dnhom}), temos
\noindent
\begin{equation}
    \dfrac{\text{d}^2}{\text{d}x^2} \psi(x) + P(x) \dfrac{\text{d}}{\text{d}x} \psi(x) 
           + Q(x) \psi(x) = 0,
\label{eq:eqDif1Dhom}
\end{equation}
que pode ser resolvida do mesmo modo que (\ref{eq:eqSchrodinger1DQ}),
tomando uma expansão em série para $P(x)$,
\noindent
\begin{equation}
    P(x) = \sum_{j=0}^{\infty} P_j \, x^j,
\label{eq:expansaoP}
\end{equation}
onde $P_n$ são os coeficientes de $P(x)$, de modo que a equação 
(\ref{eq:eqDif1Dhom}) pode ser escrita na forma
\noindent
\begin{align}
    \sum_{j=0}^{\infty} (j+2) (j+1) \, c_{j+2} \, x^j + 
         \sum_{j=0}^{\infty} \left( \sum_{i=0}^{j} (i+1) c_{i+1} 
                        P_{j-i} \right) x^j \, +& \nonumber \\
       + \sum_{j=0}^{\infty} \left( \sum_{i=0}^{j} c_i Q_{j-i} \right) x^j &= 0,
\label{eq:eqDif1DhomPQexpand2}
\end{align}
e pela unicidade, novamente, obtemos a relação de recorrência
\noindent
\begin{equation}
    \boxed{c_{j+2} = - \dfrac{1}{(j+2)(j+1)} \sum_{i=0}^{j} 
                     \bigr[ (i+1) c_{i+1} P_{j-i} + c_i Q_{j-i} \bigr] },
\label{eq:eqDif1DhomRelRec}
\end{equation}
que, dados os coeficientes das funções $P(x)$ e $Q(x)$ arbitrárias, e 
dadas as duas sementes, determina todos os coeficientes da solução de 
(\ref{eq:eqDif1Dhom}) a partir de $c_2$, notando que valem exatamente 
as mesmas considerações feitas acima para (\ref{eq:eqSchrodinger1DRelRec}) 
a respeito de sementes e soluções independentes.

Aqui vale a pena dedicar um momento de reflexão sobre a relação
(\ref{eq:eqDif1DhomRelRec}).
Essa relação de recorrência é de fato tão especial que, juntamente com as
sementes e as expressões para $P(x)$ e $Q(x)$, contém toda a informação 
necessária para resolver qualquer equação diferencial ordinária homogênea
de interesse físico com uma acurácia arbitrária e, ainda mais, permite a 
implementação de uma rotina de aritmética simples para calcular a solução
termo a termo, progressivamente um após o outro até que se atinja a 
tolerância especificada.

Por fim, para resolver a equação diferencial ordinária heterogênea 
(\ref{eq:eqDif1Dnhom}), basta tomar a expansão de $R(x)$ em série,
\noindent
\begin{equation}
    R(x) = \sum_{j=0}^{\infty} R_j \, x^j,
\label{eq:expansaoR}
\end{equation}
e, procedendo analogamente ao que foi feito para a equação homogênea, 
temos 
\noindent
\begin{equation}
    \boxed{c_{j+2} = - \dfrac{1}{(j+2)(j+1)} \left\{ \sum_{i=0}^{j} 
                     \bigr[ (i+1) c_{i+1} P_{j-i} + c_i Q_{j-i} \bigr] 
                     + R_j \right\} },
\label{eq:eqDif1DhetRelRec}
\end{equation}
que, diferentemente da relação da equação homogênea 
(\ref{eq:eqDif1DhomRelRec}), pode gerar uma solução não-trivial mesmo 
quando as duas sementes são nulas, $c_0=0$ e $c_1=0$, se os coeficientes 
$R_n$ estiverem disponíveis.
Essa terceira solução para a equação heterogênea é chamada de solução
particular \cite{ARFKEN}, e sabemos que a solução mais geral da equação 
heterogênea é dada por uma combinação linear da solução particular com
as duas soluções independentes da respectiva equação homogênea.

A equação (\ref{eq:eqDif1DhetRelRec}) sumariza diversos aspectos da 
matemática de equações diferenciais ordinárias, como a ligação entre a 
independência de soluções e a independência dos coeficientes semente, e 
ainda possibilita que se use um método prático, direto, puramente 
algébrico e iterativo para encontrar uma solução com acurácia arbitrária 
em um dado intervalo finito para qualquer equação diferencial ordinária 
homogênea ou heterogênea com $P(x)$, $Q(x)$ e $R(x)$  conhecidas e 
físicas.

A equação (\ref{eq:eqDif1DhetRelRec}) será denominada relação de 
recorrência mestre para equações diferenciais ordinárias de segundo grau.

\section{Resolvendo a equação de Schr\"odinger}
\label{secEqSchrod}

Nesta seção a relação de recorrência mestre para a equação de 
Schr\"odinger é obtida e aplicada a problemas simples em uma dimensão.

\subsection{A equação de Schr\"odinger para uma partícula em 1-D}
\label{secSchrodEq1D}

A equação de Schr\"odinger independente do tempo para uma partícula de 
massa $m$ em uma coordenada $x$ pode ser obtida da respectiva equação 
dependente do tempo nos casos em que o potencial é invariante no tempo 
usando a separação de variáveis \cite{GRIFFITHS}, o que resulta na forma
\noindent
\begin{equation}
    \hat{H} \psi(x) = E \, \psi(x),
\label{eq:1DtdepSchrodinger}
\end{equation}
onde o hamiltoniano $\hat{H}$ é dado pela soma dos operadores energia 
cinética $\hat{T}$ e potencial $\hat{V}$, $\hat{H} = \hat{T} + \hat{V}$,
o operador energia cinética para a partícula é $\hat{T} = \hat{p}^2/2m$, 
onde o operador momento linear é dado por 
$\hat{p} = - i \hbar \, \text{d}/\text{d}x$, $\hbar$ é a constante de 
Planck $h$ dividida por $2\pi$, o operador energia potencial pode ser 
substituído por uma simples função da coordenada, $\hat{V}=V(x)$, $E$ é 
o autovalor de energia, e $\psi(x)$ é a autofunção, ou função de onda, 
em uma coordenada e independente do tempo, que é um campo escalar 
complexo de quadrado integrável construído para fornecer a descrição 
mais completa possível do estado quântico de um sistema.

Isto é, para sistemas que estão nos chamados estados puros, a função de 
onda contém a informação para descrever o sistema completamente, 
tornando possível obter as medidas observáveis do cálculo dos valores 
esperados dos operadores correspondentes\footnote{
Mesmo quando o sistema não está em um estado puro, i.e., está em uma 
mistura estatística, construir a matriz densidade ainda requer as 
funções de onda dos estados puros.}.
A propriedade de quadrado integrável de $\psi(x)$ garante que a norma e 
outros valores esperados mantenham-se finitos e, portanto, físicos.
Além disso, para problemas em uma dimensão podemos tomar o campo $\psi(x)$
puramente real, sem perda de generalidade \cite{GRIFFITHS}.

A interpretação mais aceita para a função de onda é a interpretação 
estatística dada por Born, que diz que 
$|\psi(x)|^2 \, \text{d}x = \psi^*(x) \, \psi(x) \, \text{d}x$
é a probabilidade de encontrar a partícula entre $x$ e $x + \text{d}x$.
Para satisfazer a interpretação estatística, a norma quadrática da função
de onda deve, a qualquer tempo, ser igual à um\footnote{
Em situações em 
que a norma da função de onda não é conservada, e.g., em processos 
dissipativos, ainda é possível recorrer ao formalismo da matriz 
densidade, talvez especificando um reservatório acoplado ao sistema de 
interesse.},
\noindent
\begin{equation}
    \int\limits_{-\infty}^{\infty} |\psi(x)|^2 \, \text{d}x = 1.
\label{eq:PsiOrthog}
\end{equation}
É fácil ver da equação (\ref{eq:1DtdepSchrodinger}) que se $\psi(x)$ 
é uma solução complexa então o produto de $\psi(x)$ por uma constante 
complexa também é, de modo que é sempre possível normalizar uma solução, 
multiplicando-a por um número real para fazê-la obedecer à condição de 
normalização (\ref{eq:PsiOrthog}).
Além disso, ainda sobra uma fase arbitrária, e dizemos que a função de 
onda é definida exceto por uma fase global.
De posse da função de onda normalizada, torna-se possível obter o valor 
esperado de uma variável observável $\hat{A}$ tomando
\noindent
\begin{equation}
    \bigr<\hat{A}\bigr> = \int\limits_{-\infty}^{\infty} 
        \psi^*(x) \hat{A} \, \psi(x) \,\, \text{d}x,
\label{eq:AopExpectedValue}
\end{equation}
onde $\hat{A}$ é um operador hermitiano, o que faz de
$\bigr<\hat{A}\bigr>$ um valor real, e para obter o desvio padrão basta 
tomar $\sigma^2_A = \bigr<\hat{A}^2\bigr> - {\bigr<\hat{A}\bigr>}^2$.

De fato, na mecânica quântica, todos os objetos são uma onda, isto é, a 
matéria, a carga e todas as outras propriedades dessas partículas se 
comportam como ondas, são ondas de matéria e carga distribuídas no 
espaço e, como tais, apresentam todas as características naturais das 
ondas, assim também como algumas outras características muito 
particulares da mecânica quântica das ondas de matéria.
Dessa natureza ondulatória da função de onda surge o importantíssimo 
princípio de incerteza de Heisenberg, que permeia toda a natureza 
limitando a quantidade de informação que podemos extrair a respeito das 
propriedades das partículas.
Por exemplo, as variáveis posição e momento 
são quantidades canonicamente conjugadas, de modo que os operadores 
posição $\hat{x}$ e momento $\hat{p}$ não comutam, 
$[\hat{x},\hat{p}] = \hat{x}\hat{p} - \hat{p}\hat{x} = i \hbar$, o que 
obriga os desvios padrão de medidas do momento e da posição a 
obedecerem a $\sigma_x \sigma_p \geq \hbar/2$, ou seja, não é possível 
conhecer com precisão arbitrária simultaneamente o momento e a posição 
da partícula.
Isso é simplesmente uma propriedade das ondas, as funções de onda nos
espaços das coordenadas e dos momentos são pares da transformada de 
Fourier, com um fator de transformação $\hbar$, de modo que, quanto mais 
localizado o pacote de ondas estiver no espaço das coordenadas, mais ele 
estará disperso no espaço dos momentos, e vice-versa.

Uma visão esclarecedora a respeito do significado físico da função de 
onda vem do teorema de Hellmann-Feynman que, em tradução livre do artigo 
do Feynman de 1939 \cite{FEYNMAN1939}, diz:
\begin{quote}
\textit{A força em qualquer núcleo (considerado fixo) em qualquer sistema 
de núcleos e elétrons é simplesmente a força eletrostática clássica
exercida no núcleo em questão pelos outros núcleos e pelas distribuições 
de densidade de carga dos elétrons.}
\end{quote}
Essa interpretaçao é de fato tão concreta que a maioria dos métodos
modernos para cálculos quânticos, como métodos Hartree-Fock e 
pós-Hartree-Fock, interpretam a densidade $q |\psi(x)|^2$ para todos os 
efeitos práticos como sendo a própria distribuição espacial de carga de 
uma partícula de carga $q$.
E, de modo similar, $m |\psi(x)|^2$ pode ser interpretado como a 
distribuição de massa da partícula.

Agora, usando a solução polinomial (\ref{eq:seriePot}) fica fácil 
calcular a densidade de probabilidade,
\noindent
\begin{equation}
    |\psi(x)|^2 = \phi_{\infty}^*(x) \, \phi_{\infty}(x) = \sum_{j=0}^{{\infty}}
           \left( \sum_{i=0}^{j} c_i c_{j-i} \right) x^j,
\label{eq:PsiOrthogSerie}
\end{equation}
mas, quando a série é finita, as integrais em (\ref{eq:PsiOrthog}) e 
(\ref{eq:AopExpectedValue}) devem ser calculadas sobre um intervalo 
finito $[a,b]$ que contenha a vasta maioria da densidade de 
probabilidade, o intervalo da fração principal como discutido acima, e
então, para um observável qualquer, escrevemos o pseudo valor esperado
\noindent
\begin{equation}
    \bigr<\hat{A}\bigr>_{N} = \int\limits_{a}^{b} 
        \phi_N^*(x) \hat{A} \, \phi_N(x) \,\, \text{d}x,
\label{eq:AopExpectedValueFin}
\end{equation}
de modo que, no limite onde o intervalo vai a infinito, $a \to -\infty$ e 
$b \to \infty$, e simultaneamente o polinômio se torna uma série com 
$N \to \infty$, o pseudo valor esperado $\bigr<\hat{A}\bigr>_N$ se torna 
o valor esperado exato $\bigr<\hat{A}\bigr>$.

Afortunadamente, como mencionado acima, à medida que aumentamos o número 
de termos do polinômio, vai aumentando monotonicamente o tamanho do 
intervalo em que $\phi_N(x)$ aproxima a solução dentro da tolerância 
especificada, e o polinômio vai progressivamente `aderindo' à solução 
a partir de $x=0$.
Primeiro adere ao bojo central da densidade de probabilidade e depois vai 
aderindo às regiões mais distantes onde a densidade é praticamente zero e 
vai decaindo exponencialmente para ambos os lados.
Assim, sempre deve ser possível encontrar um intervalo \textit{maximin} 
que contenha a fração principal da partícula e no qual um polinômio 
\textit{minimax} aproxima muito bem as funções de onda que são soluções 
da equação de Schr\"odinger (\ref{eq:1DtdepSchrodinger}), de modo que 
todas as integrais do tipo (\ref{eq:AopExpectedValue}) podem ser 
aproximadas com acurácia arbitrária por integrais do tipo 
(\ref{eq:AopExpectedValueFin}).
Na prática, dada a expansão em série da função $Q(x)$ da equação
(\ref{eq:eqSchrodinger1DQ}), dadas duas sementes à escolha e dada uma
tolerância $\tau$ para o erro da aproximação, podemos então calcular os
coeficientes a partir de $c_2$ usando (\ref{eq:eqSchrodinger1DRelRec}), 
e a cada termo podemos determinar o intervalo \textit{maximin} do 
polinômio, notando que a função erro $\varepsilon_N(x)$ dada por 
(\ref{eq:eqDif1DhomErr}) é praticamente nula dentro desse intervalo 
e depois diverge rapidamente para fora dele, conforme discutido acima.
Então, o critério para determinar se o intervalo aproximante contém a 
fração principal da função de onda pode ser estabelecido especificando 
que a diferença relativa entre os valores esperados calculados em duas 
iterações subsequentes deve ser menor que a tolerância,
\noindent
\begin{equation}
    \left| \dfrac{\bigr<\hat{A}\bigr>_N - 
    \bigr<\hat{A}\bigr>_{N+1}}{\bigr<\hat{A}\bigr>_N} \right| < \tau.
\label{eq:critFracPrinc}
\end{equation}

O melhor da série de potências é que ela torna trivial resolver 
analiticamente as integrais das normas e dos valores esperados, obtendo 
uma expressão algébrica para (\ref{eq:AopExpectedValueFin}), de modo que 
seu cálculo numérico também deve ser praticamente imediato.
Por exemplo, o pseudo valor esperado do operador posição $\hat{x}$ 
assume a forma simples
\noindent
\begin{align}
    \bigr<\hat{x}\bigr>_{N} &= \int\limits_{a}^{b} 
        \phi_N^*(x) \, x \, \phi_N(x) \,\, \text{d}x = 
        \sum_{j=0}^{N} \left( \sum_{i=0}^{j} c_i c_{j-i} \right) 
              \int\limits_{a}^{b} x^{j+1} \,\, \text{d}x  =  \nonumber \\
        &= \sum_{j=0}^{N} \dfrac{b^{j+2}-a^{j+2}}{j+2} 
               \sum_{i=0}^{j} c_i c_{j-i},
\label{eq:AopExpectedValueFinX}
\end{align}
e de maneira análoga para todos os outros operadores 
relevantes.

Podemos obter uma boa estimativa para o intervalo que contém a fração 
principal da função de onda observando a convergência da solução no 
infinito.
Suponhamos que o comportamento assintótico é um decaimento exponencial, 
como acontece com os estados ligados do poço quadrado de potencial 
finito, de modo que 
$\psi(x) \sim C \text{e}^{- \alpha |x|}$ para $x \to \pm \infty$, onde 
$\alpha > 0$.
A fração do valor esperado que cai fora do intervalo da fração principal, 
ou seja, a fração residual do valor esperado, é dada por
\noindent
\begin{equation}
    \varepsilon_{res} = 
    \int\limits_{-\infty}^{a} \psi^*(x) \hat{A} \, \psi(x) \,\, \text{d}x \, + \,
    \int\limits_{b}^{\infty} \psi^*(x) \hat{A} \, \psi(x) \,\, \text{d}x.
\label{eq:AopExpectedValueSec}
\end{equation}
Supondo, por simplicidade, que o intervalo é simétrico, $a=b$, que o 
valor esperado é a norma, ou seja, o operador é a identidade $\hat{I}$, 
e supondo também que as constantes de decaimento para a direita e para 
a esquerda são idênticas, então a fração residual fica
\noindent
\begin{equation}
    \varepsilon_{res} = 
    2 C^2 \int\limits_{b}^{\infty} \text{e}^{- 2 \alpha x} \,\, \text{d}x = 
    \dfrac{C^2}{\alpha} \text{e}^{- 2 \alpha b}.
\label{eq:AopExpectedValueSecSim}
\end{equation}
Assim, o erro no cálculo da norma cometido ao excluir a fração residual
descresce exponencialmente com o tamanho do intervalo $2b$, e podemos 
usar (\ref{eq:AopExpectedValueSecSim}) com um critério de tolerância como 
$| \varepsilon_{res} | \leq \tau$ para obter uma expressão direta para a
estimativa do tamanho do intervalo da fração principal,
\noindent
\begin{equation}
    2 b \geq - \text{ln} (\alpha \tau / C^2 ) / \alpha,
\label{eq:intervOk}
\end{equation}
considerando obviamente que a tolerância deve ser $\tau < C^2/ \alpha$
para que o logarítmo resulte em um valor negativo, e o tamanho do 
intervalo seja um valor positivo.
Se tomamos outro operador, por exemplo o operador energia cinética 
$\hat{T} = - (\hbar^2/2m) \, \text{d}^2/\text{d}x^2$, o valor da fração 
residual em (\ref{eq:AopExpectedValueSecSim}) será multiplicado por um 
fator $\alpha^2 \hbar^2 / 2m$, de modo que a estimativa 
(\ref{eq:intervOk}) é uma boa estimativa inicial para o intervalo da
fração principal no cálculo do valor esperado de qualquer operador,
lembrando que, dado o problema concreto, sempre é possível estabelecer 
um bom intervalo para cada operador individualmente, ou simplesmente um 
intervalo que seja extenso o suficiente para aproximar bem todos os 
operadores.

Então, usando a série de Maclaurin para aproximar o comportamento de 
decaimento exponencial das funções de onda de interesse, o intervalo de 
domínio de cada termo do polinômio aumenta linearmente com $N$ e, ainda, 
a componente da fração residual dos valores esperados decai 
exponencialmente com o tamanho do intervalo.
Assim, deve ser possível encontrar uma solução acuradíssima para um vasto 
intervalo da coordenada e, mesmo que a série contenha dezenas, centenas 
de milhares de coeficientes, deve ser possível calcular qualquer valor 
esperado de modo praticamente imediato nos computadores modernos.

Agora, fazendo as substituições competentes em 
(\ref{eq:1DtdepSchrodinger}), obtemos
\noindent
\begin{equation}
    - \dfrac{\hbar^2}{2m} \dfrac{\text{d}^2}{\text{d}x^2} \psi(x) + V(x) \psi(x) = 
        E \psi(x),
\label{eq:eqSchrodinger1D}
\end{equation}
que é equivalente a tomarmos a equação (\ref{eq:eqSchrodinger1DQ}) com a 
função
\noindent
\begin{equation}
    Q(x) = \dfrac{2m}{\hbar^2} \bigr( E - V(x) \bigr),
\label{eq:eqQschrod1D}
\end{equation}
e, se expandirmos $V(x)$ em série de potências,
\noindent
\begin{equation}
    V(x) = \sum_{j=0}^{\infty} V_j \, x^j,
\label{eq:expansaoV}
\end{equation}
pela unicidade das séries temos que os coeficientes da função $Q(x)$ são
dados diretamente por $Q_0=2m(E - V_0)/\hbar^2$ e $Q_j=-2m V_j/\hbar^2$ 
para $j>0$, levando à relação de recorrência mestre para a equação de
Schr\"odinger,
\noindent
\begin{equation}
    \boxed{c_{j+2} = \dfrac{1}{(j+2)(j+1) } \dfrac{2m}{\hbar^2} \left\{
                     \sum_{i=0}^{j} c_i V_{j-i} - c_j E \right\} }.
\label{eq:eqSchrod1DRelRec}
\end{equation}

A função $Q(x)$ pode ser identificada com o quadrado do número de onda 
$k$ da função de onda no ponto $x$. 
Mais especificamente, a relação entre o momento da partícula $p$ e as 
propriedades da onda associada é dada pela relação de de Broglie,
\noindent
\begin{equation}
    p=hk=\frac{h}{\lambda}.
\label{eq:deBroglie}
\end{equation}
Ou seja, quando a energia potencial varia no espaço então o momento da 
partícula e consequentemente seu comprimento de onda $\lambda$ também 
variam.
Em regiões do espaço onde a energia potencial é baixa, a energia cinética 
é alta e o momento linear é alto, portanto o comprimento de onda é baixo, 
e vice-versa.
Em uma configuração espacial de potencial com características múltiplas,
não necessariamente contínuas, a frequência da função de onda muda de 
modo suave para acompanhar cada característica particular.

Quanto ao potencial $V(x)$, este é em geral conhecido, podemos usar 
termos coulombianos, acoplamento \textit{spin}-órbita, correções 
relativísticas, entre outros de acordo com as características do 
problema mas, infelizmente, não conhecemos \textit{a priori} a função 
$Q(x)$ porque a energia $E$ é uma constante global desconhecida, de modo
que nem sempre é possível usar a equação (\ref{eq:eqSchrod1DRelRec})
diretamente.
É necessário, então, analisar mais detalhadamente o problema de se 
encontrar os autovalores de energia.

Para uma partícula em mecânica quântica há sempre duas possibilidades, 
ou a partícula é livre ou está em um estado ligado.
Podemos fazer a distinção mais claramente definindo \textit{prima facie}
um referencial absoluto de energia considerando que, em todo problema
realista de mecânica quântica, o potencial deve ir a zero no infinito.
O espaço muito além do sistema de estudo deve ser sempre livre para todos 
os fins práticos, ou seja, sem qualquer potencial, como acontece em
qualquer sistema realista de átomos e moléculas.
Com essa convenção, temos diretamente que a partícula é livre quando 
$E>0$, e quando $E<0$ a partícula está ligada a algum tipo de poço de 
potencial, por exemplo, um elétron no campo elétrico produzido por um 
núcleo atômico, ou um átomo preso em um potencial anarmônico de uma rede 
cristalina.
Importa notar que, para um estado ligado, na região espacial em que a 
energia da partícula é maior que a energia potencial, a solução 
apresenta comportamento oscilatório e é chamada de região classicamente 
permitida e, por outro lado, na região em que a energia da partícula é 
menor que a energia potencial, a densidade de probabilidade decai a zero
exponencialmente até o infinito, e é chamada região classicamente 
proibida.

Agora, se a partícula é livre, então a energia pode ser qualquer dentro 
de um contínuo de valores, o usuário pode escolher livremente $E$ de 
acordo com o objetivo de sua pesquisa, por exemplo um estudo de 
espalhamento, e então usar a equação (\ref{eq:eqSchrod1DRelRec}) para 
obter a solução direta e com acurácia arbitrária.

Por outro lado, se a partícula está em um estado ligado, a solução 
somente tem caráter físico para determinados valores discretos de $E$, e 
se a energia  escolhida não for um desses valores quantizados permitidos 
então a solução obtida de (\ref{eq:eqSchrod1DRelRec}) não será física.
Mais especificamente, para os estados ligados, podemos saber se uma dada
energia leva a uma solução física analisando a solução no infinito, 
equação (\ref{eq:limPsiInf}), se ela não vai a zero no infinito então ela
não  é física \cite{ARFKEN}.

Assim, quando se trata de encontrar os autovalores discretos de uma 
partícula ligada a um poço de potencial, tendo em mente a condição 
(\ref{eq:limPsiInf}), então o problema agora pode ser formulado da 
seguinte maneira: encontrar o intervalo \textit{maximim} que contenha 
toda a fração principal da função de onda, que para todos os efeitos 
numéricos torna os pseudo valores esperados em valores esperados, e no 
qual um polinômio \textit{minimax} aproxima a solução dentro da 
tolerância especificada, e em uma configuração de energia que minimiza a 
função de onda nos extremos do intervalo.

Dentro do esquema que está sendo apresentado, podemos pensar em uma 
diversidade de métodos para encontrar todos os autovalores discretos de 
energia para um dado potencial, apesar dessa informação não estar 
explicitamente codificada na equação de Schr\"odinger.
Por exemplo, uma possibilidade é tomar um grau $N$ do polinômio bem alto, 
de modo que o intervalo de aproximação seja bem amplo, e então varrer os 
valores de energia desde a energia mínima do poço até a energia da sua 
borda mais baixa, calculando (\ref{eq:eqSchrod1DRelRec}) e buscando pelos 
valores de $E$ que minimizam a solução nos extremos do intervalo.
Esse método é bastante simples de ser implementado e deve funcionar 
de modo muito eficiente, ao menos para problemas em uma dimensão.
Contudo, como veremos no capítulo seguinte, há um método ainda mais
simples e mais direto de obter uma excelente estimativa para todos os 
autovalores de qualquer problema envolvendo um ou múltiplos poços de
potencial, e depois ainda é possível refinar esses autovalores a uma
acurácia arbitrária.

Cabe aqui nesta seção um último comentário de ordem prática, sobre o 
cálculo numérico da série de potências.
Os coeficientes da expansão em série de uma função com decaimento 
exponencial como $\text{e}^{- \alpha x}$ são proporcionais a $1/j!$ e,
portanto, rapidamente se tornam números muito pequenos com o aumento do 
índice, o que pode acabar gerando \textit{underflow} numérico se a 
precisão do tipo é fixa\footnote{
Uma alternativa é usar rotinas de precisão arbitrária, mas essas são 
mais lentas que as rotinas usuais, e provavelmente seria muito difícil 
usá-las com um bom método de minimização numérica.}.
Por exemplo, com o tipo numérico de ponto flutuante de precisão dupla
(64 \textit{bits}), podemos representar números com expoentes de até 308, 
de modo que somente são representáveis os fatoriais de 170 ou menores.
Para evitar o \textit{underflow} numérico no cálculo da expansão de 
$\text{e}^{- \alpha x}$, basta rearranjar os termos da expansão de 
seguinte maneira
\noindent
\begin{equation}
    \text{e}^{- \alpha x} = \sum_{j=0}^{\infty} \dfrac{{(- \alpha x)}^j}{j!} =
        1 + \sum_{j=1}^{\infty} \, \prod_{i=1}^{j} \dfrac{{- \alpha x}}{i},
\label{eq:serieExp}
\end{equation}
de modo que, calculando o produtos das frações ${- \alpha x}/i$, a 
magnitude dos termos intermediários fica sempre muito menor do que 
calculando ${(- \alpha x)}^j$ e dividindo por $j!$.
Agora, quando precisamos tratar de fato com coeficientes muito pequenos, 
por exemplo, em uma rotina de minimização numérica, uma maneira simples 
é escrever o coeficiente como um produto de dois coeficientes, por 
exemplo, $2 \cdot 10^{-600} = 2 \cdot 10^{-300} \cdot 10^{-300}$.
De maneira mais geral, qualquer coeficiente $c_j$ pode ser escrito como 
uma produtória de coeficientes componentes, $c_j = \prod_{i=1}^{j} c'_{ji}$, 
de modo que sempre é possível calcular a série em alguma forma de 
produtórias do tipo (\ref{eq:serieExp}) sem gerar \textit{underflow}.
Mesmo quando a produtória dos coeficientes não contém o mesmo número de 
termos que $x^j$ é sempre possível combinar ou dividir coeficientes de 
modo que ambas as produtórias possuam o mesmo número de termos, e então 
a divisão entre as produtórias pode ser escrita como a produtória da 
divisão dos termos como na forma (\ref{eq:serieExp}), mantendo portanto 
sob controle a magnitude dos termos intermediários.
A função de onda é, em todos os casos, uma função contínua e limitada, 
oscilando em certa região do espaço e passando de maneira suave a um 
comportamento exponencial decrescente, ainda que o potencial apresente 
descontinuidades, de modo que nada de fato diverge.

Agora, sabendo que a função de onda apresenta, por exemplo, decaimento 
exponencial decrescente em $x>0$, podemos escrever a solução polinomial 
(\ref{eq:seriePot}) de maneira mais bem informada fatorando esse 
comportamento exponencial, por exemplo, aplicando a transformação 
$c_j = b_j \, (- \alpha)^j / j!$,
\noindent
\begin{equation}
    \phi_N(x) = \sum_{j=0}^{N} c_j \, x^j = 
        \sum_{j=0}^{N} \, \dfrac{b_j {(- \alpha x)}^j}{j!} =
        b_0 + \sum_{j=1}^{N} \, b_j \prod_{i=1}^{j} \dfrac{{- \alpha x}}{i},
\label{eq:seriePotFator}
\end{equation}
de modo que o valor de $c_j$, que tende a ficar muito pequeno com $j$, é 
multiplicado por um outro valor, $j! / (- \alpha)^j$, que tende a ficar 
muito grande.
Na prática, o valor de $\alpha$ usado na transformação de coeficientes 
pode ser estimado calculando a razão entre dois coeficientes da série 
para índices altos, já que no decaimento exponencial cada coeficiente é 
proporcional ao anterior por um fator $\alpha/j$, mais especificamente, 
$\alpha \sim -(j+1) c_{j+1}/c_j$ para $j \gg 1$.
O valor de $\alpha$ não precisa ser o valor exato do coeficiente real de 
decaimento, de fato, usar o valor exato em (\ref{eq:seriePotFator}) é a 
melhor fatoração, mas quando $\alpha$ é somente uma boa estimativa da 
constante de decaimento, ainda assim deve ser possível encontrar uma boa 
fatoração, ao menos uma que retarde o \textit{underflow} numérico em 
muitos e muitos coeficientes da série.
Além disso, nesse esquema de fatoramento, a constante $C$ pode ser 
facilmente obtida dos coeficientes fatorados $b_j$, já que a razão entre
dois coeficientes subsequentes $b_{j+1}/b_j$ tende a um, e então 
$b_j \to C$ para $j \to \infty$.
O fato mais importante é que sempre podemos construir algum tipo de 
fatoração de comportamento que torna a série mais enxuta, e inclusive é
possível trabalhar com mais de um fatoramento por solução, por exemplo,
podemos manter uma fatoração para $x \ll 0$ e outra para $x \gg 0$, 
enquanto o centro da distribuição usa os coeficientes não fatorados 
$c_j$.
E, além do mais, a fatoração também pode ser aprimorada à medida que se 
calcula os coeficientes da série, e se torna possível determinar com 
maior acurácia a constante de decaimento.

Nas seções seguintes será demonstrada a aplicação direta da relação de 
recorrência mestre da equação de Schr\"odinger (\ref{eq:eqSchrod1DRelRec}) 
para alguns problemas simples e de solução bem conhecida na mecânica 
quântica, para ilustrar o conceito da solução por série de potências.
Estes são os problemas arquetípicos em mecânica quântica em uma dimensão
para os quais a energia exata dos estados pode ser obtida trivialmente,
de modo que é possível usar a relação de recorrência mestre diretamente.

No capítulo seguinte trataremos de problemas um pouco mais complicados,
onde inicialmente podemos contar somente com uma boa estimativa da 
energia dos estados, sendo necessário aplicar um processo de refinamento 
posterior dos autovalores de energia que utilizará como guia a relação 
de recorrência mestre.

\subsection{O poço quadrado de potencial}
\label{secOscLinear}

Nesta seção, para efeito de demonstração, a relação de recorrência mestre 
para a equação de Schr\"odinger (\ref{eq:eqSchrod1DRelRec}) é aplicada de 
modo direto para resolver os problemas dos poços quadrados de potencial 
finito e infinito, levando às conhecidas soluções senóides e exponenciais
\cite{GRIFFITHS}.
As derivações apresentadas nesta seção e nas seguintes, através da 
construção de uma solução a partir da relação de recorrência mestre para 
a equação de Schr\"odinger, são derivações simples e fundamentais e, como 
tal, tem também seu interesse educacional.

O poço quadrado infinito é uma idealização do problema onde as paredes 
são perfeitamente reflexivas e, por convenção, para o poço simetricamente 
centrado na origem e de largura $2a$, tomamos o potencial 
%
\noindent
\begin{equation}
  V(x) = \left\{ 
    \begin{array}{l l}
       0,             & \quad \text{para} \,\, |x| < a, \\
       \infty,        & \quad \text{para} \,\, |x| > a,
    \end{array} \right.
\label{eq:infPotWellV}
\end{equation}
de modo que, em todos os casos, $E>0$, e há somente estados ligados.
Assim, nessa versão idealizada, a função de onda é nula fora do poço e 
oscilatória dentro dele, uma combinação linear de seno e cosseno, sendo 
que as energias permitidas são obtidas impondo as condições de contorno 
nas bordas do poço onde a função de onda deve ser nula, $\psi(\pm a)=0$.
Como rotina, o número de onda dentro do poço é definido por 
$k \equiv \sqrt{2mE} / \hbar$, e a condição de contorno impõe que a 
extensão $2a$ do poço seja um múltiplo inteiro de $\lambda/2$, de modo
que $k_n = n \pi / 2 a$, onde o inteiro $n \ge 1$ é o número quântico,
que diz quantas excitações há nos modos de vibração da partícula dentro
do poço.
A energia pode então ser diretamente obtida combinando essas condições,
$E_n = \hbar^2 k_n^2 / 2m$, e é proporcional ao quadrado do número de 
excitações.

Agora, tomando $V(x)=0$ em (\ref{eq:eqSchrod1DRelRec}) temos que 
\noindent
\begin{equation}
    c_{j+2} = - \dfrac{c_j}{(j+2)(j+1)} \dfrac{2mE_n}{\hbar^2} = 
              - \dfrac{c_j k_n^2}{(j+2)(j+1)},
\label{eq:eqSchrod1DRelRecInfWell}
\end{equation}
e, redefinindo as sementes em termos de duas constantes reais $A=c_0$ 
e $B=c_1/k_n$, obtemos a solução mais completa na forma
\noindent
\begin{equation}
    \psi(x) = A \left[1 - \dfrac{(k_n x)^2}{2!} + \dfrac{(k_n x)^4}{4!} - \cdots \right] +
              B \left[k_n x - \dfrac{(k_n x)^3}{3!} + \dfrac{(k_n x)^5}{5!} - \cdots \right],
\label{eq:seriePotInfWell1}
\end{equation}
que pode ser expressa simplesmente como combinação linear de seno e 
cosseno,
\noindent
\begin{equation}
    \psi(x) = A \, \text{cos}(k_n x) + B \, \text{sin}(k_n x),
\label{eq:seriePotInfWell2}
\end{equation}
de modo que fazendo $A=0$ obtemos a solução anti-simétrica, que atende à 
condição de contorno $\psi(\pm a)=0$ para $n$ par, e fazendo $B=0$ 
obtemos a solução simétrica, que atende à condição de contorno para $n$ 
ímpar.
Em ambos os casos, as constantes de normalização $A$ e $B$ são facilmente 
calculadas da condição de normalização da função de onda 
(\ref{eq:PsiOrthog}), resultando em $a^{-1/2}$.

No caso do poço quadrado finito, simetricamente centrado na origem, de 
largura $2a$ e profundidade $V_0$, a função potencial é dada por 
%
\noindent
\begin{equation}
  V(x) = \left\{ 
    \begin{array}{l l}
       -V_0,         & \quad \text{para} \,\, |x| < a, \\
       0,            & \quad \text{para} \,\, |x| > a,
    \end{array} \right.
\label{eq:finPotWellV}
\end{equation}
onde $V_0>0$ e, para uma partícula ligada, $0>E>-V_0$, de modo que na
região $|x| < a$ a solução é oscilatória, e na região $|x| > a$ a solução
é exponencial decaindo a zero no infinito para ambos os lados.
Neste caso, a solução tradicional é, assim como o potencial, construída 
por partes e portanto não é analítica, e as energias permitidas são
obtidas impondo-se a continuidade da solução e sua primeira derivada nas 
bordas do poço $x=\pm a$.
Dentro do poço finito, $E>V(x)$ e define-se o número de onda como 
$k \equiv \sqrt{2m(E+V_0)} / \hbar$, e então as soluções são obtidas do
mesmo modo que (\ref{eq:seriePotInfWell2}).
Fora do poço, $E<V(x)=0$ e define-se o número de onda como 
$\kappa \equiv \sqrt{-2mE} / \hbar$.

Agora, tomando $V(x)=0$ em (\ref{eq:eqSchrod1DRelRec}), e notando que
agora $-E$ é um valor positivo, temos que para $|x|>a$,
\noindent
\begin{equation}
    c_{j+2} = \dfrac{c_j}{(j+2)(j+1)} \dfrac{2m|E_n|}{\hbar^2} = 
              \dfrac{c_j \kappa_n^2}{(j+2)(j+1)},
\label{eq:eqSchrod1DRelRecFinWell}
\end{equation}
e, novamente, redefinindo as sementes em termos de duas constantes reais 
$A=c_0$ e $B=c_1/\kappa_n$, obtemos a solução mais completa na forma
\noindent
\begin{equation}
    \psi(x) = A \left[1 + \dfrac{(\kappa_n x)^2}{2!} + \dfrac{(\kappa_n x)^4}{4!} + \cdots \right] +
              B \left[\kappa_n x + \dfrac{(\kappa_n x)^3}{3!} + \dfrac{(\kappa_n x)^5}{5!} + \cdots \right],
\label{eq:seriePotFinWell1}
\end{equation}
que, por exemplo, tomando $A=B$, nos dá a solução exponencial crescente 
para $x<-a$
\noindent
\begin{equation}
    \psi(x) = A \left[1 + \kappa_n x + \dfrac{(\kappa_n x)^2}{2!} + \dfrac{(\kappa_n x)^3}{3!} + \cdots \right]
            = A \, \text{e}^{\kappa_n x},
\label{eq:seriePotFinWellPos}
\end{equation}
e, tomando $A=-B$, nos dá a solução exponencial decrescente para $x>a$
\noindent
\begin{equation}
    \psi(x) = A \left[1 - \kappa_n x + \dfrac{(\kappa_n x)^2}{2!} - \dfrac{(\kappa_n x)^3}{3!} + \cdots \right]
            = A \, \text{e}^{-\kappa_n x}.
\label{eq:seriePotFinWellNeg}
\end{equation}

Os autovalores do problema do poço quadrado finito de potencial são mais 
difíceis de se obter que os autovalores do poço quadrado infinito, 
pois parte da função de onda 'vaza' para fora do poço, ou seja, parte da 
energia da onda tunela através da parede do poço, de modo que o 
comprimento da onda dentro do poço sofre uma pequena dilatação que pode 
chegar até $\pi/2$.
Para os estados de menor energia, o decaimento exponencial na região 
classicamente proibida é forte porque $\kappa$ é alto, então a solução
oscilatória dentro do poço tem seu comprimento de onda ligeiramente
dilatado para fazer com que a inclinação da onda ao atravessar a borda do 
poço seja justamente a mesma dada pelo início do decaimento fortemente 
exponencial.
Para os estados de energia mais alta, $\kappa$ é menor e o decaimento
exponencial é mais suave, há uma maior penetração da densidade de 
probabilidade da partícula na região classicamente proibida, e portanto 
a onda dentro do poço perde mais energia, e seu comprimento de onda tem 
que ser mais fortemente dilatado para atender à inclinação desse 
decaimento exponencial mais suave ao atravessar a borda do poço.
Os autovalores podem ser encontrados impondo a continuidade da função
de onda e sua derivada na borda do poço levando às bem conhecidas 
equações transcendentais para a energia, 
$k_n = \kappa_n \, \text{tan}(\kappa_n a)$ para a solução simétrica, e 
$k_n = \kappa_n \, \text{cot}(\kappa_n a)$ para a solução anti-simétrica.
Do gráfico dessas equações \cite{GRIFFITHS} é facil notar que, para os
estados de menor energia, $\kappa$ é muito próximo de um múltiplo de 
$\pi/2a$, ou seja, $\kappa \sim n \pi/2a$, ficando ligeiramente abaixo 
desse valor e, então, com o aumento do número de excitações $n$, a 
diferença de $\kappa$ para $n \pi/2a$ vai aumentando, até o limite do 
estado ligado de maior energia do poço onde $\kappa$ pode perder no 
máximo $\pi/2a$, ou seja, $\kappa \sim (n-1) \pi/2a$, ficando na prática 
sempre um pouco acima desse valor.

\subsection{O potencial linear}
\label{secPotLinear}

Nesta seção a relação de recorrência mestre para a equação de 
Schr\"odinger é aplicada para resolver diretamente o problema do 
potencial linear, levando à forma exata conhecida, as funções de Airy 
\cite{GRIFFITHS}.

Um potencial linear, por exemplo aquele experimentado por uma partícula
de carga $q$ em um campo elétrico externo constante $\mathcal{E}/q$, pode 
ser escrito na forma
\noindent
\begin{equation}
    V(x) = \mathcal{E} x,
\label{eq:linPotV}
\end{equation}
de modo que $Q_1=\mathcal{E}$ e $Q_j=0$ para $j \ne 1$.
Notando que neste problema idealizado o potencial não tem estados 
ligados, que portanto a energia pode assumir qualquer valor dentro de um 
contínuo de valores permitidos e dado que, neste caso em particular, um 
deslocamento de energia significa simplesmente que a função de onda é 
deslocada linearmente na coordenada sem deformação, então podemos tomar 
$E=0$ sem perda de generalidade.
Com essa convenção, temos que, para $\mathcal{E}>0$, a solução deve ser 
oscilatória na região $x<0$, e na região $x>0$ deve decair a zero no 
infinito mais rápido que a exponencial, e vice-versa para 
$\mathcal{E}<0$.

Agora, calculando (\ref{eq:eqSchrod1DRelRec}) para $j=0$ temos que 
$c_2=0$ pois $Q_0=0$.
Já para $j \ge 1$, definindo\footnote{
O procedimento é o mesmo para $\mathcal{E}<0$.}
$\mathcal{E}>0$, tomando o potencial $\mathcal{E} x$ em 
(\ref{eq:eqSchrod1DRelRec}), e com $E=0$, temos que
\noindent
\begin{equation}
    c_{j+2} = \dfrac{c_{j-1} }{(j+2)(j+1)} \dfrac{2m \mathcal{E}}{\hbar^2} = 
              \dfrac{c_{j-1} \alpha^3}{(j+2)(j+1)},
\label{eq:eqSchrod1DRelRecLinPot}
\end{equation}
onde $\alpha = (2m \mathcal{E}/\hbar^2)^{1/3}$ é um número positivo.
Para $j=1$ temos que $c_3=\alpha^3 c_0 / (3 \cdot 2)$, para $j=2$ temos
$c_4=\alpha^4 c_1 / (4 \cdot 3)$, para $j=3$ temos $c_5=0$ já que 
$c_2=0$, para $j=4$ temos que 
$c_6=\alpha^6 c_0 / (6 \cdot 5 \cdot 3 \cdot 2)$, e assim por diante, de 
modo que, redefinindo as sementes em termos de duas constantes reais 
$A=c_0$ e $B=c_1 / \alpha$, obtemos a solução mais completa na forma
\noindent
\begin{equation}
    \psi(x) = A \left[1 + \dfrac{(\alpha x)^3}{3 \cdot 2} + 
                          \dfrac{(\alpha x)^6}{6 \cdot 5 \cdot 3 \cdot 2} + \cdots \right] +
              B \left[\alpha x + \dfrac{(\alpha x)^4}{4 \cdot 3} + 
                          \dfrac{(\alpha x)^7}{7 \cdot 6 \cdot 4 \cdot 3} + \cdots \right],
\label{eq:serieLinPot1}
\end{equation}
que é justamente a forma das funções de Airy Ai e Bi \cite{BENDER},
\noindent
\begin{equation}
    \psi(x) = A \, \text{Ai}(\alpha^{1/3} x) + B \, \text{Bi}(\alpha^{1/3} x).
\label{eq:serieLinPot2}
\end{equation}
Sabemos que a função Bi diverge em $x \to \infty$, então a escolha 
particular das sementes que leva a uma solução física neste caso é dada
por $A \ne 0$ e $B = 0$.

%% file: bibli.tex
\renewcommand{\bibname}{Referências}

%% file: paperQuant.bbl
\begin{thebibliography}{999}

 
\bibitem{SCHRODINGER} E. Schr\"odinger, 
\textit{An Undulatory Theory of the Mechanics of Atoms and Molecules.}
The Physical Review, \textbf{28} (1926), 1049--1070.

\bibitem{NOBEL}
\textit{The official web site of the nobel prize.}
<http://nobelprize.org>, página acessada em 09/08/2014.

\bibitem{NATUREMAT} 
\textit{Crossing length scales.} Editorial,
Nature Materials, \textbf{12} (2013), 1079.

\bibitem{NATURECHEM} X. Deupi, 
\textit{Molecular dynamics: A stitch in time.}
Nature Chemistry, \textbf{6} (2014), 7--8.

\bibitem{ARFKEN} G. B. Arfken, H. J. Weber,
\textit{Mathematical Methods for Physicists.} 5$^{\rm nd}$ edition,
Harcourt/Academic Press (2001).

\bibitem{GRIFFITHS} D. J. Griffiths,
\textit{Introduction to quantum mechanics.}
Prentice Hall, New Jersey (1995).

\bibitem{COHEN} C. Cohen-Tannoudji, B. Diu, F. Laloe,
\textit{Quantum Mechanics.}
Hermann e John Wiley \& Sons, Paris, França (1992).

\bibitem{TWOPHOTON} D. M. Greenberger, M. Horne, A. Zeilinger,
\textit{Bell theorem without inequalities for two particles. I. Efficient detectors.}
Physical Review A, \textbf{78} (2008), 022110.

\bibitem{MINIMAX} E. W. Weisstein,
\textit{Minimax Polynomial. From MathWorld -- A Wolfram Web Resource.}
<http://mathworld.wolfram.com/MinimaxPolynomial.html>, página 
acessada em 06/08/2014.

\bibitem{BENDER} C. M. Bender, S. A. Orszag,
\textit{Advanced mathematical methods for scientists and engineers.}
McGraw-Hill (1978).

\bibitem{HYLLERAAS1} M. B. Ruiz,
\textit{Hylleraas Method for Many-Electron Atoms. I. The Hamiltonian.}
International Journal of Quantum Chemistry, \textbf{101} (2005), 246--260.

\bibitem{HYLLERAAS2} J. Rychlewski, J. Komasa,
\textit{Explicitly correlated functions in variational calculations.}
Em J. Rychlewski (ed.), 
\textit{Explicitly correlated wave functions in chemistry and physics. Theory and applications.}
Kluwer Academic Publishers, Dordrecht, Países Baixos (2003).

\bibitem{FEYNMAN1939} R. P. Feynman,
\textit{Forces in Molecules.}
Physical Review, \textbf{56} (1939), 340--343.




\end{thebibliography}
